\documentclass[aps,prc,reprint,noeprint,superscriptaddress]{revtex4-2}

\usepackage[T1]{fontenc}
\usepackage[varg]{newtx}
\usepackage[colorinlistoftodos]{todonotes}
\usepackage[normalem]{ulem}
\usepackage{graphicx}
\usepackage{xcolor}
\usepackage{multirow}
\usepackage[colorlinks,citecolor=blue]{hyperref}
\usepackage{orcidlink}
\graphicspath{{./}{figures/}}

\newcommand{\rpn}{\textit{r}~process}
\newcommand{\rpa}{\textit{r}-process}
\newcommand{\eqc}[1]{Eq.~\eqref{#1}}

\newcommand{\subrf}[2]{Fig.~\hyperref[#1]{\ref{#1}(#2)}}

\newcommand{\dzsh}{FRDM$^*$}
\newcommand{\frdmsh}{DZ31$^*$}

\definecolor{crimson}{RGB}{220,20,60}

\begin{document}


\title{Impact of nuclear masses on $r$-process nucleosynthesis:
bulk properties versus shell effects}


\author{Samuel~A.~Giuliani\,\orcidlink{0000-0002-9814-0719}}
\affiliation{Departamento de F{\'i}sica Te{\'o}rica and CIAFF, Universidad
Aut{\'o}noma de Madrid, Madrid 28049, Spain}

\author{Gabriel~Mart{\'i}nez-Pinedo\,\orcidlink{0000-0002-3825-0131}}
\affiliation{GSI Helmholtzzentrum f{\"u}r Schwerionenforschung,
  Planckstra{\ss}e 1, D-64291 Darmstadt, Germany}
\affiliation{Institut für Kernphysik (Theoriezentrum),  Fachbereich Physik,
  Technische Universität Darmstadt, Schlossgartenstra{\ss}e 2, D-64289
  Darmstadt, Germany}
\affiliation{Helmholtz Forschungsakademie Hessen f\"ur FAIR (HFHF),
  GSI Helmholtzzentrum f\"ur Schwerionenforschung, Planckstra{\ss}e~1,
  D-64291 Darmstadt, Germany}

\author{Andreas Bauswein\,\orcidlink{0000-0001-6798-3572}}
\affiliation{GSI Helmholtzzentrum f{\"u}r Schwerionenforschung,
  Planckstra{\ss}e 1, D-64291 Darmstadt, Germany}
\affiliation{Helmholtz Forschungsakademie Hessen f\"ur FAIR (HFHF),
  GSI Helmholtzzentrum f\"ur Schwerionenforschung, Planckstra{\ss}e~1,
  D-64291 Darmstadt, Germany}

\author{Vimal Vijayan\,\orcidlink{0000-0002-4690-2515}}
\affiliation{GSI Helmholtzzentrum f{\"u}r Schwerionenforschung,
  Planckstra{\ss}e 1, D-64291 Darmstadt, Germany}


\date{\today}

\begin{abstract}
We investigate the impact of the model estimating the masses of exotic nuclei
on \rpa\ nucleosynthesis, assessing the dependence of the abundance
distribution on the specific properties of nuclear masses.  By decomposing
theoretical nuclear mass predictions into a liquid-drop parametrization and
local shell effects, we show that \rpa\ abundances are virtually insensitive to
large variations of the masses which originate from nuclear bulk properties of
the model, such as the symmetry energy. In contrast, the mass component
associated with local shell effects is the main driver of \rpa\ abundance
variations, despite its relatively minor contribution to the absolute value of
neutron separation energies.  Our work suggests that experimental and theoretical studies
of masses devoted to \rpa\ applications, such as the nucleosynthesis in the
ejecta of neutron star mergers, should focus on the physical origin and
determination of local changes in mass trends.
\end{abstract}


\maketitle

\section{Introduction}
The cosmic origin of heavy elements is a
major challenge in modern science~\cite{NRC2003}. Roughly half of the
elements heavier than iron are produced in a nucleosynthesis
process known as rapid neutron-capture process, or
\rpn~\cite{Burbidge1957,Cameron1957}. It occurs in an astrophysical
environment with extreme neutron fluxes, where exotic neutron-rich
nuclei are produced by subsequent neutron captures and $\beta$
decays. As such, a fundamental understanding of the \rpn\ requires a
multidisciplinary effort involving the modeling of astrophysical
environments, astronomical observations as well as the knowledge of
nuclear and atomic properties of the synthesized nuclei 
(see~\cite{Horowitz2018,Kajino2019,Arnould2020,Cowan2021} for recent reviews).

When it comes to the modeling of \rpa\ abundances, the two major
sources of uncertainty are the astrophysical conditions at which the
\rpn\ operates and the nuclear properties of the involved nuclei. In
recent years, increasing efforts have been made to
understand which astrophysical scenarios could provide suitable
conditions for the production of heavy elements and identify
observational constraints. Numerical simulations suggest three main
candidates: compact binary mergers, such as binary neutron star
mergers (NSM)~\cite{Symbalisty1982} and NS-black hole
mergers~\cite{Lattimer1974}; magneto-rotational core-collapse
supernova~\cite{Symbalisty1985}; and accretion disk outflows from
collapsars~\cite{MacFadyen1999}. So far, the only direct evidence of
the production of heavy elements has been obtained for NSM, through
the observation of the AT2017gfo kilonova light
curve~\cite{Abbott2017b,Villar2017}, the electromagnetic
counterpart associated to the gravitational wave event
GW170817~\cite{Abbott2017a}. 

From the nuclear physics side, the main challenge resides in the fact
that the nuclei produced during the \rpn\ are extremely neutron rich
and short-lived. As such, measuring the relevant reaction rates in
current experimental facilities is often not possible, and one must
rely on theoretical modeling that results in large differences in
predictions as the neutron drip-line is approached. Uncertainties in
the nuclear physics input and astrophysical conditions produce
degeneracies in the predicted abundances, hindering the identification
of the astrophysical conditions and relevant nuclear properties by
confronting \rpa\ models and observations.  Among the nuclear
properties that are required for modeling the \rpn, nuclear mass
is the most basic quantity, as it determines the energy threshold of
all reactions and decays.
As a consequence, several experimental campaigns in radioactive
ion-beam facilities have been devoted to the measurement of masses of
neutron-rich nuclei, and more experiments are currently ongoing and
envisaged to extend the reach further into the \rpa\
region~\cite{Horowitz2018,Kajino2019,Cowan2021,Brown.Gade.ea:2024}. At
the same time, more refined global nuclear models capable of predicting
nuclear masses across the entire nuclear chart have been developed, some of
them capable of reproducing experimental data with a root-mean-square (rms) error
smaller than
700~keV~\cite{Duflo1995,Goriely2010,Liu2011,Moeller.Sierk.ea:2016,Grams.Shchechilin.ea:2024}.
In recent years, these global calculations of nuclear masses have been
supplemented with machine learning techniques, to alleviate the
computational burden and/or effectively reduce model discrepancies with
experimental
data~\cite{Utama2017,Niu2018a,Neufcourt2020,Lasseri2020,Mumpower2023,Saito2023}.

Despite all the progress made, discrepancies among models in the predicted
masses tend to increase with neutron excess. Because of the fundamental role
that nuclear masses play for the \rpn, it has always been assumed that absolute
differences have \emph{per se} a major impact on \rpa\ abundances. Therefore,
experimental efforts have been dedicated to constraining the value of nuclear
masses with the highest possible precision~\cite{Horowitz2018}.  From a theory
side, several studies explored the impact of different mass models on predicted
\rpa\
abundances~\cite{Arcones2011,Arcones2012,Mendoza-Temis2015,Mumpower2015,Martin2016a,Sprouse2020,Vassh2021,Zhu2021,Kullmann2023a}.
However, when comparing different theoretical predictions (among themselves or
against experimental data), there has been so far no attempt to determine which
features in the evolution of nuclear masses with neutron and proton number, the
so-called mass surface, are important and which ones are negligible for the
determination of the \rpa\ abundances.

In this work, we show that variations in the predicted masses that are
related to differences in their bulk or global properties, e.g.\ symmetry
energy, remarkably, do not affect the \rpa\ abundances. Instead, the
nucleosynthesis outcome is highly sensitive to shell effects that induce local
changes in neutron separation and neutron shell gap energies. We decompose the
predicted masses into two contributions: one that represents the bulk properties
and changes smoothly with proton and neutron number, and a second part that
accounts for local variations on top of the smooth global behavior. The first
part is described using a liquid drop model parametrization obtained from a fit
to a specific mass model, while the second represents the difference. We
consider two mass models which have been widely used in \rpa\ nucleosynthesis
studies: the Finite-Range Droplet Model (FRDM)~\cite{Moller1995} and the
Duflo-Zuker mass formula (DZ31)~\cite{Duflo1995}.
We then generate new mass tables by mixing the smooth energy part and shell
corrections of the original models. By performing nuclear network
calculations for a large set of trajectories corresponding to the dynamical
ejecta of a neutron star merger~\cite{Collins.Bauswein.ea:2023}, 
we demonstrate the insensitivity of predicted abundances to large
variations of the nuclear masses, provided that they originate from the smooth
part of the underlying nuclear mass model. In contrast, we show that local
shell effects are the main drivers in shaping the abundances distribution.

The paper is organized as follows. In Sec.~\ref{sec:decomposition} we describe
the methodology employed for decomposing nuclear masses into a smooth bulk part
and local shell effects. In Sec.~\ref{sec:results} we present the abundances
predicted by mass models differing in their bulk properties and/or shell effects.
Our findings and conclusions are summarized in Sec.~\ref{sec:conclusions}.

\section{Decomposition of nuclear masses\label{sec:decomposition}} 
We assume that the binding energy of a nucleus can
be decomposed into an average contribution, which depends smoothly on the
number of protons ($Z$) and neutrons ($N$), and a quantum shell-correction
arising from local changes in single-particle levels~\cite{Strutinsky1967}. The
homogeneous, bulk part of the energy can be efficiently parametrized using a
leptodermous expansion such as the liquid drop model (LDM)~\cite{Myers1982}:
	\begin{align}
		 \label{eq:LDM} \frac{E^\textup{LDM}}{A} & =
		a_\textup{vol} + a_\textup{sur} A^{-1/3} + a_\textup{cur}
		A^{-2/3} 
		 + a_\textup{sym} I^2 \\ &+ a_\textup{ssym} A^{-1/3} I^{2}  + a^{(2)}_\textup{sym} I^4 
		 + a_\textup{Coul} Z^2 A^{-4/3} + a_\textup{pai} \delta\,,
       \nonumber
	\end{align}
being $A=N+Z$, and $I=(N-Z)/A$ the so-called neutron excess. The pairing term
$\delta$ takes the value $A^{-3/2}$ for even-even nuclei, $-A^{-3/2}$ for
odd-odd nuclei, and zero for odd-mass nuclei.  A detailed discussion of the
physical meaning of the different terms entering in \eqc{eq:LDM} can be found,
for instance, in Refs.~\cite{Myers1982,Reinhard2006}. Starting from a mass
table (either DZ31 or FRDM), we fit the $a_i$ coefficients of \eqc{eq:LDM} to
parametrize the smooth behavior of mass surfaces. The residuals between nuclear
masses and the smooth contribution constitute the shell effects of the original
mass model.  By combining the homogeneous contribution obtained from the DZ31
(FRDM) mass table with the shell effects of FRDM (DZ31), we create a new mass
table denoted \frdmsh (\dzsh). Table~\ref{tab:ldm_par} summarizes the LDM
parameters of the mass tables employed in this work.
\begin{table}
	\centering
	\begin{ruledtabular}\begin{tabular}{lcccccccc}
		Model & $a_\textup{vol}$ & $a_\textup{sur}$ & $a_\textup{cur}$	
		& $a_\textup{sym}$ & $a_\textup{ssym}$ & $a^{(2)}_\textup{sym}$
		& $a_\textup{Coul}$ & $a_\textup{pai}$ \\ 
		\hline
		DZ31 & \multirow{2}{*}{$-$14.96} & \multirow{2}{*}{13.16} &
		\multirow{2}{*}{7.96} & \multirow{2}{*}{27.91} &
		\multirow{2}{*}{$-$27.97} & \multirow{2}{*}{$-$3.22} &
		\multirow{2}{*}{0.67} & \multirow{2}{*}{$-$9.05} \\
		\frdmsh\ & & & & & & & & \\
		\hline
		FRDM & \multirow{2}{*}{$-$14.56} & \multirow{2}{*}{11.21} &
		\multirow{2}{*}{10.39} & \multirow{2}{*}{26.16} &
		\multirow{2}{*}{$-$23.10} & \multirow{2}{*}{$-$1.28} &
		\multirow{2}{*}{0.66} & \multirow{2}{*}{$-$9.86} \\
		\dzsh\ & & & & & & & & \\
        \hline
        AME20 & $-$15.63 & 19.17 & $-$4.74 & 26.48 & $-$20.56 & $-$5.21 & 0.68 & $-$6.78
	\end{tabular}\end{ruledtabular}
	\caption{LDM parameters (in MeV) of \eqc{eq:LDM} corresponding to the
          mass models employed in this work. The last row corresponds to the LDM parameter 
          obtained from the AME2020 experimental masses~\cite{Wang2021}.\label{tab:ldm_par}}
\end{table}

Evidently, the four mass tables predict rather different nuclear masses,
particularly in the region far from stability, where the original models DZ31
and FRDM have not been fitted. This is shown in \subrf{fig:masses}{a}, where
nuclear masses along the neodymium isotopic chain are plotted as a function of
neutron number $N$. 
\begin{figure}[tb]
	\includegraphics[width=0.9\columnwidth]{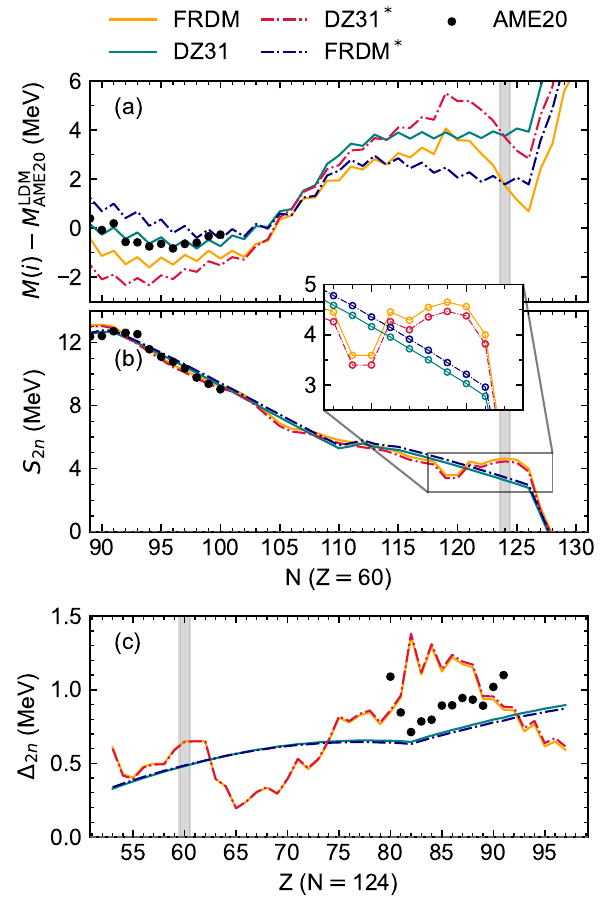}%
	\caption{Panel (a): Comparison of theoretical and experimental masses
		for neodymium ($Z=60$) isotopes as a function of neutron number. The bulk LDM 
        AME2020 contribution
        is subtracted from all the masses.
		Panel (b): Comparison of two-neutron separation energies. Panel
		(c): comparison of two-neutron shell-gap energies as a function of
		charge number for $N=124$ isotones. In all the panels, black
		circles represent AME2020 experimental
	data.\label{fig:masses}}
\end{figure}
By subtracting the homogeneous contribution given by the AME2020 LDM
parametrization (obtained from the fit of Eq.~\eqref{eq:LDM} to the AME2020
experimental masses~\cite{Wang2021}, see last row in Table~\ref{tab:ldm_par}),
it becomes visible that the model pairs DZ31/\dzsh\ and FRDM/\frdmsh\ predict
different values of nuclear masses, while sharing the same local shell effects.
Due to this, despite the large discrepancies in the binding energies, we
observe that the neutron separation energies $S_{2n}$ predicted by the
different models show a rather similar behavior, as depicted in
\subrf{fig:masses}{b} (for displaying purposes, we plot $S_{2n}$ rather than
$S_{n}$).  The modified mass tables result in a slight change in the $S_{2n}$
slope, mostly due to changes in the symmetry energy term $a_\textup{sym}$,
which governs the evolution of masses with neutron excess. We recall that
neutron-capture rates depend exponentially on the neutron separation energy,
suggesting that the impact of variations in nuclear masses can be easily
overestimated, as local changes in $S_{2n}$ are more relevant for shaping the
\rpa\ abundances. Such local changes can be quantified by means of the
(two-)neutron shell-gap energies, which are given by the difference between
two-neutron separation energies $\Delta_{2n} (N,Z) = S_{2n} (N,Z) - S_{2n}
(N+2,Z)$.  This quantity is extremely sensitive to variations in
single-particle levels between neighboring nuclei, providing a proxy for local
changes in the binding energies due to nucleonic (shell)
effects~\cite{Satua1998,Dobaczewski2001a,Bengtsson1981b}.  For instance, local
maxima in $\Delta_{2n}$ are associated with the presence of spherical and
deformed shell gaps, while negative values usually indicate a nuclear shape
transition (see Ref.~\cite{Buskirk2023} for a recent detailed discussion on
$\Delta_{2n}$ and its evolution across the nuclear chart). The $\Delta_{2n}$
values along the isotonic chain $N=124$ predicted by the four models are
plotted in \subrf{fig:masses}{c}. As one can see, this quantity is insensitive
to changes in the bulk properties of nuclear masses characterized by the LDM
parametrization, as its evolution is actually driven by the emergence and
disappearance of shell effects with neutron excess.

\section{Results\label{sec:results}}
Nucleosynthesis calculations are performed using a
large nuclear reaction network, including all nuclei up to $Z=110$ from the
valley of stability to the neutron dripline. We derived neutron capture rates
consistently for each mass model by means of the Hauser-Feshbach statistical
theory, employing the TALYS~1.95 nuclear reaction code~\cite{Koning2007}.
Experimental masses are employed whenever available. Photodissociation rates
are obtained from neutron-capture rates using detailed balance. We adopted the
FRDM $\beta$-decay rates, which have been renormalized for each mass table
according to their predicted $Q$-values. To isolate the impact of
masses on predicted abundances, in all the calculations we employed the
neutron-induced fission rates based on the FRDM+TF model~\cite{Panov2010}, and
the fission yields derived using the code ABLA~\cite{Kelic2009}.

We compute the \rpa\ abundances for a set of 2015 trajectories simulating the dynamical
ejecta produced in neutron star mergers~\cite{Collins.Bauswein.ea:2023}
(see~\cite{Ardevol2019} for code details).  We use this specific \rpa\ scenario
because it covers a broad range of astrophysical conditions, in particular
regarding the proton-to-nucleon ratio (see Fig.~2 of
Ref.~\cite{Collins.Bauswein.ea:2023}). By this, we show that our result is
fully general and does not depend on considering a specific trajectory. We also
expect our conclusions to be valid for any \rpa\ scenario.

\subsection{Impact of bulk properties on \rpa\ abundances} 
\begin{figure}[tb]
	\includegraphics[width=\columnwidth]{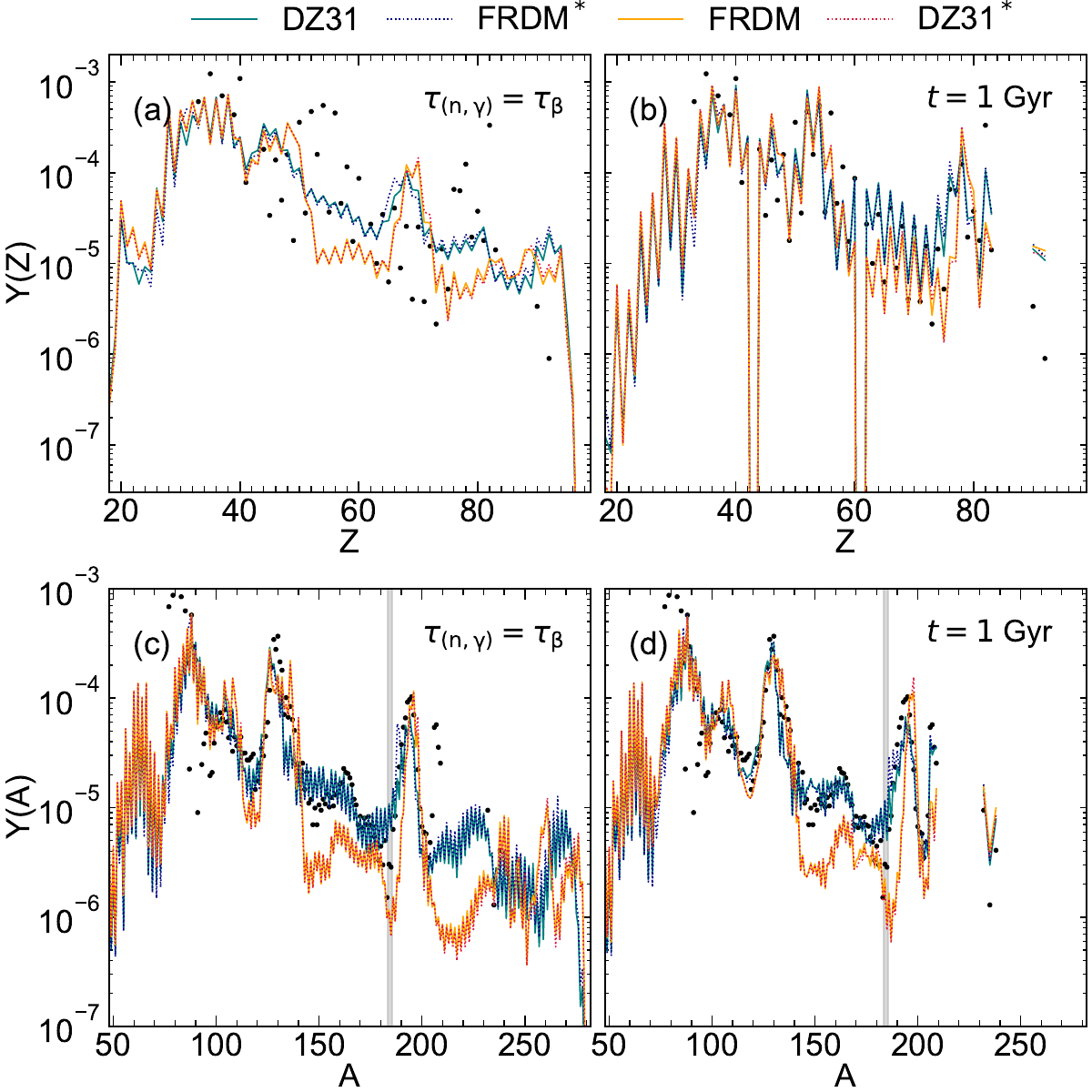}%
	\caption{Mass-integrated abundances as a function of atomic number (top
		panels) and mass number $A$ (bottom panels) predicted by
		different mass models at $\tau_{(n, \gamma)} = \tau_\beta$ (left panels) and at 1~Gyr
		(right panels). Black circles are \rpa\ abundances in the solar
		system.\label{fig:YZYA}}
\end{figure}

We start by exploring the sensitivity of \rpa\ abundances on bulk properties of
nuclear masses. Figure~\ref{fig:YZYA} shows the total integrated elemental and
mass abundances predicted from the four mass tables of Table~\ref{tab:ldm_par}
at two different phases of the evolution: when the average timescale for
neutron captures becomes equal to the average timescale for $\beta$ decay
($\tau_{(n, \gamma)} = \tau_\beta$), and at 1 Gyr.
The abundances obtained with the four models at 1~Gyr (when most of the
material has decayed to stability) present the main characteristics of a strong
\rpa\ pattern: we observe the presence of the second and third peaks, as well
as the production of uranium and thorium. When comparing the
different models, we notice that already at the freeze-out the abundances
predicted by masses with the same shell effects but different bulk properties,
are virtually identical: only minor differences in the strength of odd-even
staggering are observed in mass abundances, which are subsequently washed out
by $\beta$-delayed neutron emissions and late neutron captures. We observe no
connection between the bulk properties of masses (as determined by LDM
parameters) and the abundance distribution. In other words, models with the
same LDM coefficients $a_i$ result in very different abundances patterns. 

To better understand the fundamental connection between variations in
nuclear masses and abundances, we study the case of neodymium isotopes ($Z=60$)
around neutron number $N=124$, which dominate the abundances at freeze-out in
the region $A=184$ (indicated with a gray vertical band in
Fig.~\ref{fig:YZYA}). Figure~\hyperref[fig:YZYA]{\ref{fig:YZYA}(c)} shows that
the abundances predicted by FRDM/\frdmsh\ strikingly differ from DZ31/\dzsh. In
particular, FRDM and \frdmsh\ show a deeper trough right before the third \rpa\
peak compared to DZ31 and \dzsh. This behavior of abundance distributions
cannot be explained by the differences in the predicted masses around the
region $Z/N=60/124$. As one can notice in \subrf{fig:masses}{a}, the \frdmsh\
predictions at $N=124$ (gray band) are closer to DZ31, while those predicted by
\dzsh\ around this mass number are more in agreement with FRDM predictions.
Instead, one has to look at the predicted $S_{2n}$ shown in
\subrf{fig:masses}{b}.  Despite the absolute difference in predicted masses,
which is a consequence of variations in the bulk part of the binding energy,
FRDM and \frdmsh\ (as well as DZ31 and \dzsh) show the same $S_{2n}$ trend and,
in turn, a strikingly similar \rpa\ path. In particular, FRDM and \frdmsh\
predict a local increase in $S_{2n}$ around $Z/N=60/124$, resulting in a
larger $\Delta_{2n}$.  This change in the slope of $S_{2n}$ reduces the
probability of neutron captures for nuclei in this region, producing in turn a
deeper trough in the abundance distribution around $A \sim
184$~\cite{Arcones2011}. More generally, this result shows that features of the
\rpa\ abundances are not connected to global variations in nuclear masses, but
rather the result of local changes in binding energies as a consequence of
configuration changes in single particle levels. 

\begin{figure}
  \centering
  \includegraphics[width=0.9\columnwidth]{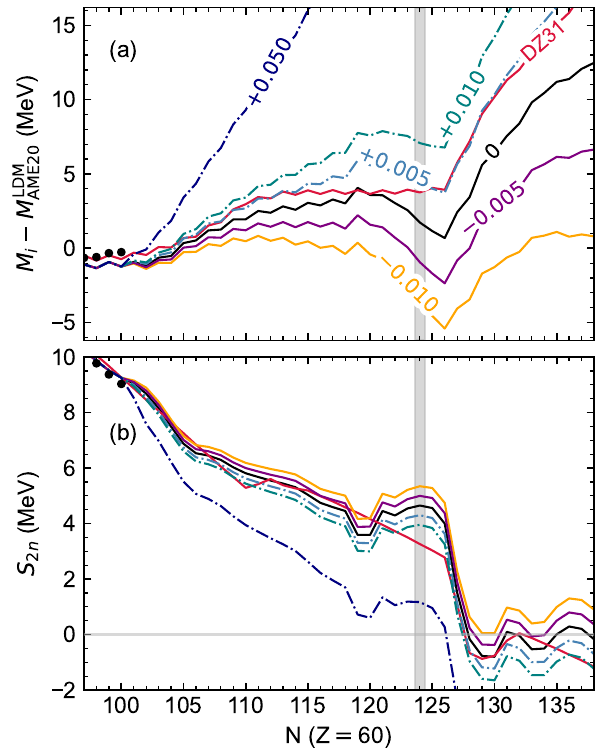}
  \caption{FRDM masses (panel (a)) and two-neutron separation energies (panel
	  (b)) in MeV for different values of $a_\textup{Nsym}$ along the
	  neodymium isotopic chain as a function of neutron number. The red
	  solid line represents the DZ31 predictions. Black circles represent
	  the AME2020 experimental data. In panel (a), the bulk LDM AME2020
	  contribution is subtracted from all the masses.\label{fig:ansym}}
\end{figure}

To further assess the impact produced by global changes of mass
surfaces, we study the variation of elemental abundances with increasing
deviations between nuclear masses.  To do so, we introduce a new term in the
LDM formula~\eqref{eq:LDM}:
\begin{equation}
  a_\textup{Nsym} [N-N_\textrm{max}^\textrm{exp}(Z)] H(N-N_\textrm{max}^\textrm{exp}(Z)) I^2 \,, 
  \label{eq:ansym}
\end{equation}
with $H(x)$ being the Heaviside step function and
$N_\textrm{max}^\textrm{exp}(Z)$ the maximum $N$ value for each isotopic chain
$Z$ with a measured mass in the AME2020 dataset~\cite{Wang2021}. By varying the
$a_\textup{Nsym}$ term, we produce mass tables that diverge at different rates
with increasing isospin asymmetry. Figure~\ref{fig:ansym} shows the nuclear
masses along the neodymium isotopic chain for five different values of
$a_\textup{Nsym}$ applied to the FRDM LDM parametrization. One can notice that
the smallest absolute values of $a_\textup{Nsym}$ roughly match the absolute
differences between FRDM and DZ31 masses, while $a_\textup{Nsym}=\pm0.01$ and
$0.05$~MeV produce mass tables that largely differ from the original FRDM
predictions.

\begin{figure*}
	\centering
	\includegraphics[width=0.9\textwidth]{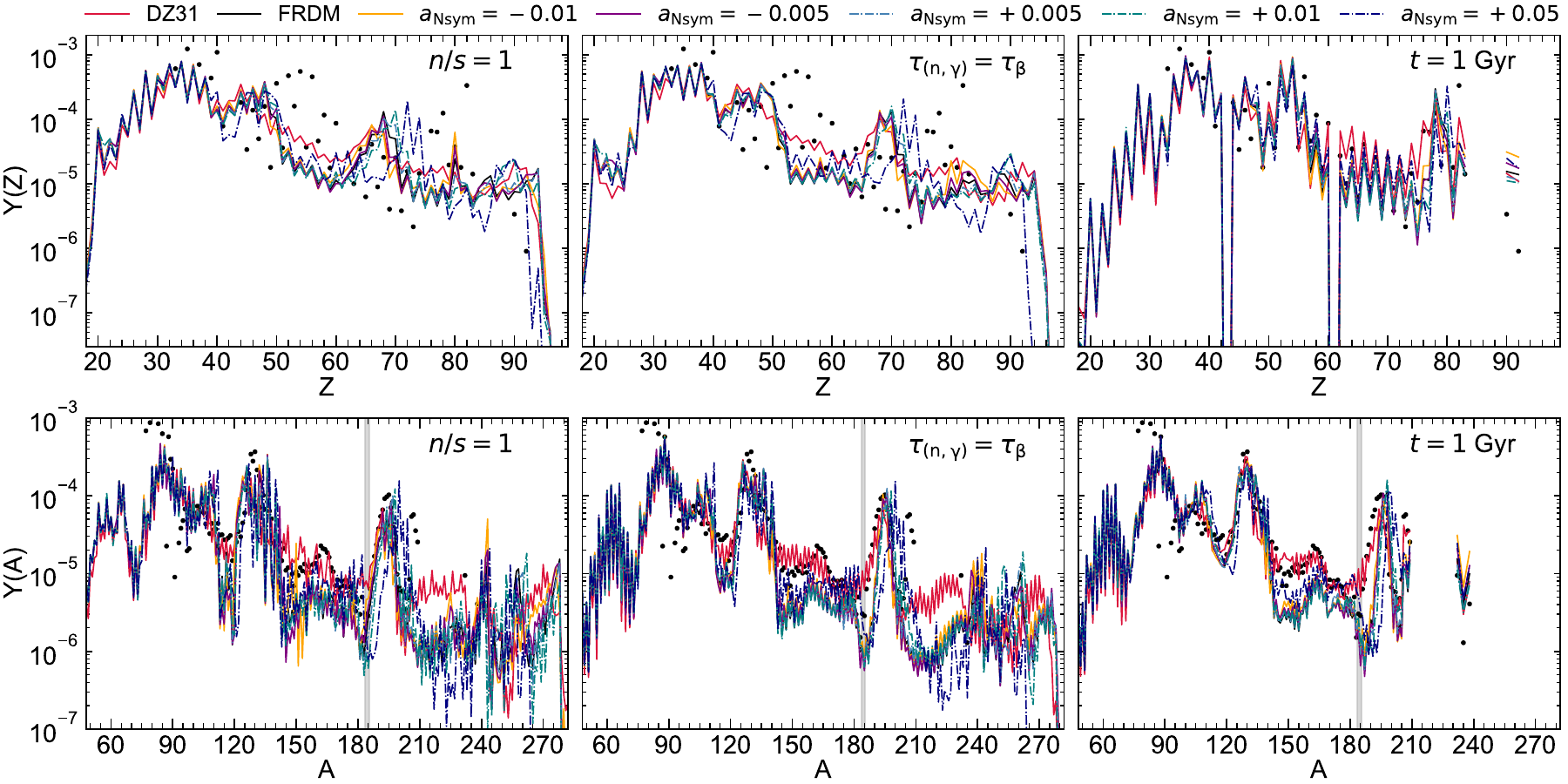}
	\caption{Integrated abundances as a function of atomic number
		$Z$ (top row)
		and mass number $A$ (bottom row) predicted by FRDM mass models
		with different values of $a_\textup{Nsym}$ and DZ31 at three
		different times: neutron
		capture freeze-out (left column), when $\tau_{(n, \gamma)} =
		\tau_\beta$
		(middle column), and at 1~Gyr (right column). Black circles are
		\rpa\ abundances in the solar system.\label{fig:YZYA_an}
		}
\end{figure*}

The integrated \rpa\ abundances obtained with these modified FRDM mass tables are shown in
Fig.~\ref{fig:YZYA_an} at three different times: at the neutron capture
freeze-out (defined as the time when the neutron-to-seed ratio $n/s=1$), 
when $\tau_{(n, \gamma)} = \tau_\beta$, and at 1~Gyr.
At early times ($n/s=1$), we observe that variations in the bulk properties of
mass surfaces produce a shift in the abundance distribution, while the overall shape
resulting from local peaks and troughs remains mostly unaltered.
As the evolution proceeds and the \rpa\ path moves closer to stability, 
the differences are washed out, producing virtually identical abundance
distributions at $\tau_{(n, \gamma)} = \tau_\beta$ and 1~Gyr.
Actually, noticeable differences in the final abundances only appear for
$a_\textup{Nsym}=0.05$~MeV in the $Z=77-81$ region. These differences are
produced by the strong decrease of the two-neutron separation energies around
the $N=126$ shell closure, with $S_{2n}$ dropping below zero for nuclei with $Z
\lesssim 68$, which shifts the third peak towards heavier atomic numbers. The
impact of variations in masses is also visible in the final-integrated
abundances as a function of mass number $A$, with a shift in the position of
troughs and peaks for $a_\textup{Nsym}=0.05$~MeV.  However, it is important to
notice that variations in the abundances between the FRDM models are generally
smaller than the variations observed between DZ31 and any FRDM-based model,
regardless of the level of agreement between the absolute masses predicted by
the different models.  This result provides further evidence that the main
drivers of \rpa\ abundances are the local changes in $S_{2n}$, rather than the
global changes of mass surfaces, suggesting that sensitivity studies focused on
individual masses could overemphasize their impact on \rpa\ abundances.

\subsection{Impact of local shell effects on \rpa\ abundances}
\begin{figure}
  \centering
  \includegraphics[width=0.9\columnwidth]{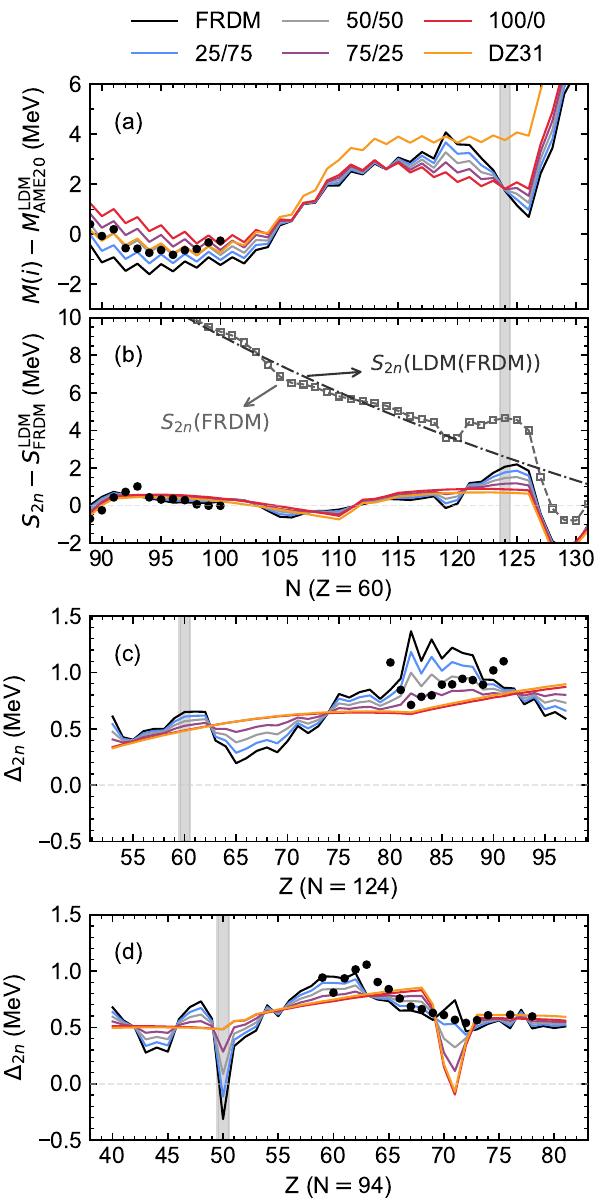}%
  \caption{Panels (a) and (b): Nuclear masses and two-neutron separation
	  energies (in MeV) for different mixtures of DZ31 and FRDM shell
	  effects (DZ31/FRDM) along the neodymium isotopic chain as a function of neutron
	  number. The bulk LDM contribution extracted from AME2020 experimental
	  data is subtracted from all the masses. The bulk LDM extracted from
	  the FRDM model is subtracted from all the $S_{2n}$. Panel (c) and (d):
	  two-neutron shell-gap energies for the same models as in the top
	  panels along the $N=124$ and $N=94$ isotonic chains as a function of proton number.
    Black circles represent the AME2020 experimental
	  data.\label{fig:mix_masses}}
\end{figure}

In the previous subsection, we showed that variations in the bulk properties of
nuclear masses have little impact on \rpa\ abundances, suggesting that variations in shell effects 
drive the changes in the abundances. The goal of this section is to
demonstrate this connection explicitly. 
To do so, we construct four mass tables with the
same LDM parametrization (derived from the FRDM model), and different shell
effects obtained by mixing the shell effects from DZ31 and FRDM in different proportions
($\delta(\text{DZ31})/\delta(\text{FRDM})=25/75$, $50/50$, $75/25$, and
$100/0$, with $\delta(i) = M(i) - \text{LDM}[M(i)]$ being the shell effects of the
mass model $i$).
The advantage of this
method compared to, for example, a general scaling of the shell effects is
that neutron shell closures are preserved, while allowing for a smooth transition
between the two mass models.
The resulting masses and two-neutron separation energies along the neodymium isotopic chain
are depicted in Figure~\ref{fig:mix_masses}, together with the
two-neutron shell gaps along the $N=124$ and 94 isotonic chains.
The two-neutron separation energies (\subrf{fig:mix_masses}{b}) and the
two-neutron shell gaps (\subrf{fig:mix_masses}{c} and \subrf{fig:mix_masses}{d}) show that, indeed, the local trends 
of the mixed mass tables gradually transition from one model to the other. 

To quantify the impact in nuclear masses induced by variations in the shell
effects, we computed the
rms error ($\sigma$) between the mixed mass models and the original FRDM\@. The results
are summarized in Table~\ref{tab:rms_mix}.
\begin{table}
	\centering
    \begin{ruledtabular}\begin{tabular}{lccccc}
                                         & 25/75 & 50/50 & 75/25 & 100/0 & DZ31 \\
    \hline
    $\sigma_\textrm{FRDM}(M)$            & 0.57  & 1.13  & 1.70  & 2.26  & 3.30 \\
    $\sigma_\textrm{FRDM}(S_{2n})$       & 0.24  & 0.48  & 0.71  & 0.95  & 0.98 \\
    $\sigma_\textrm{FRDM}(\Delta_{2n})$  & 0.27  & 0.54  & 0.82  & 1.09  & 1.09 \\[0.6ex]
    $\langle \Delta Y(Z) \rangle$ & 30.2\% & 58.0\% & 82.4\% & 105.0\% & 105.2\% \\
    $\langle \Delta Y(A) \rangle$ & 34.6\% & 68.1\% & 101.9\% & 134.41\% & 135.44\% \\[0.5ex]
    $\sigma_\textrm{AME20}(M)$           & 0.84  & 0.69  & 0.69  & 0.83  & 0.43
    \end{tabular}\end{ruledtabular}

	\caption{Deviations between mixed models and FRDM. First three rows: rms error (in MeV) for nuclear masses $M$, two-neutron separation energies $S_{2n}$, and two-neutron shell gaps $\Delta_{2n}$. Fourth and fifth rows: average deviation of elemental and mass abundance percent difference with respect to FRDM ($\Delta Y = (Y - Y_\textrm{FRDM})/Y_\textrm{FRDM} \times 100$) depicted in Fig.~\ref{fig:YZYA_mix}. 
     As a reference, the last row shows the rms errors (in MeV) for nuclear masses between the mixed mass models and AME20, while the last column summarizes the results for DZ31.\label{tab:rms_mix}}
\end{table}
Obviously, the rms errors are proportional to the percentage of DZ31 shell effects
employed in the mixed mass models: increasing the contribution of DZ31 shell
effects by 25\% produces an increase of $0.57$, $0.24$, and $0.27$ MeV in the
rms error of binding energies, two-neutron separation energies, and two-neutron
shell gaps, respectively. We point out that the rms error in $\Delta_{2n}$ 
between DZ31 and FRDM is the same as between 100\% DZ31
and FRDM, indicating that this quantity is mostly sensitive to local shell
effects, as shown in \subrf{fig:mix_masses}{c}. As a reference, Table~\ref{tab:rms_mix} 
indicates the rms error between the mixed mass models and the AME20 experimental data.
As one can see, the mixed models have a performance between FRDM ($\sigma_\textrm{AME20}=1.07\,$MeV)
and DZ31 ($\sigma_\textrm{AME20}=0.43\,$MeV).

\begin{figure*}
	\centering
	\includegraphics[width=0.9\textwidth]{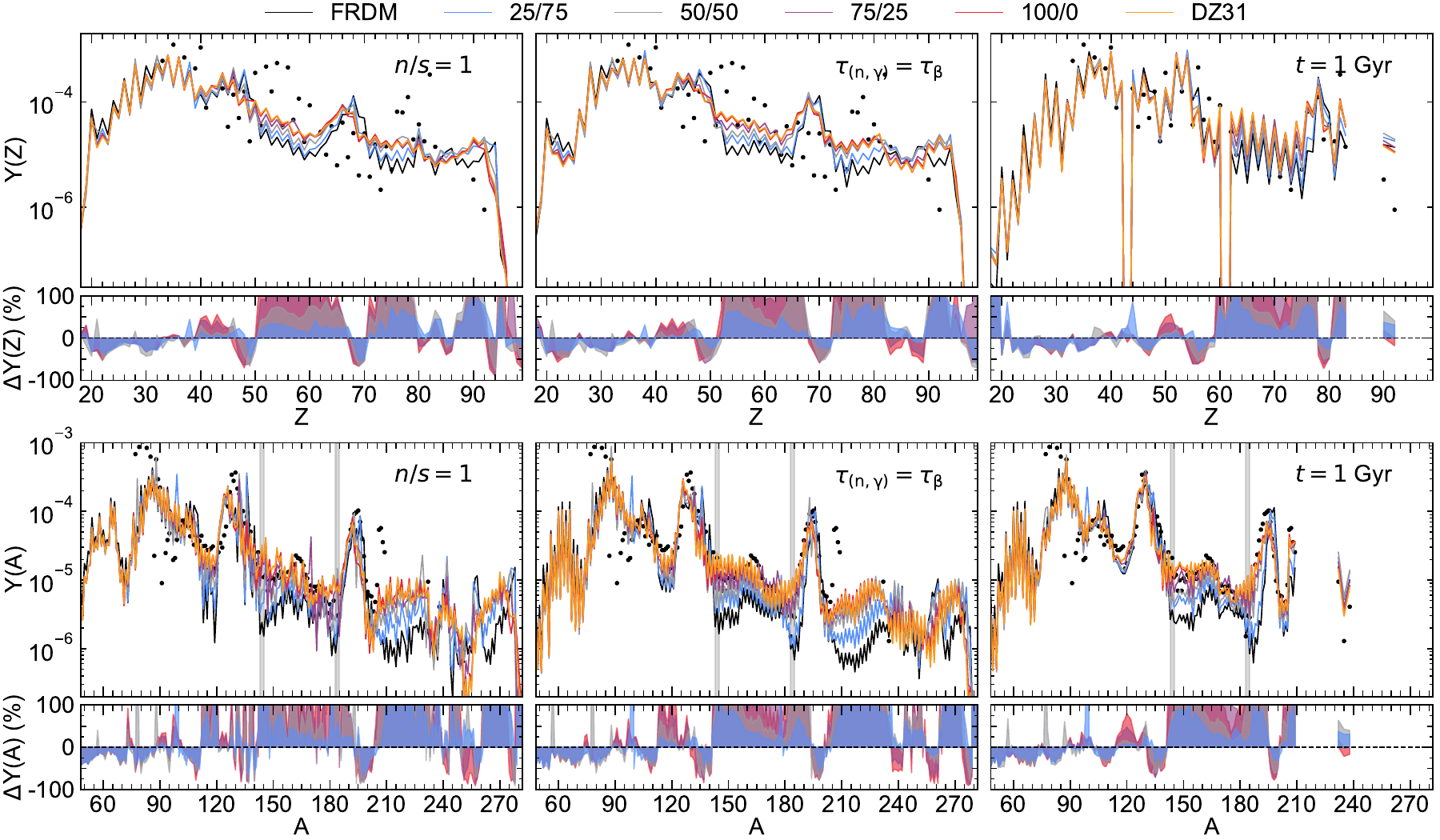}
	\caption{Integrated abundances predicted by mass models with mixed
		shell effectsc DZ31/FRDM (Table~\ref{tab:rms_mix}) at three different
		times: when $n/s=1$ (left column), when $\tau_{(n, \gamma)} =
		\tau_\beta$ (center column) and at 1~Gyr (right column).  First
		row: integrated elemental abundances as a function of atomic number $Z$.
		Second row: percent difference of integrated elemental
		abundances to FRDM\@ ($\Delta Y = (Y - Y_\textrm{FRDM}) /
Y_\text{FRDM} \times 100$). Third row: integrated mass abundances as a
		function of mass number $A$. Fourth row: percent difference of
		integrated mass abundances with respect to FRDM\@. Black circles are the
		solar \rpa\ abundances.\label{fig:YZYA_mix}}
\end{figure*}

Fig.~\ref{fig:YZYA_mix} shows the integrated elemental and mass abundances
predicted by the different mixed mass models from Table~\ref{tab:rms_mix}
at three different times
($n/s=1$, $\tau_{(n, \gamma)} = \tau_\beta$, and 1~Gyr).
For each time, we also plot the percent difference of the abundances with
respect to the FRDM results ($\Delta Y = 100 \times (Y_i - Y_\textrm{FRDM}) /
Y_\text{FRDM}$, with $i$ being one of the mixed mass models).
The smooth transition in the local shell effects in the mixed mass models introduced above 
translates into a gradual change in the abundance distributions. In
particular, the shift in the location of peaks and the progressive filling of
the troughs are clearly visible at all times.
For example, the gradual disappearance of the $S_{2n}$ 
saddle point around $Z/N=60/124$ (Figure~\ref{fig:mix_masses}) occurring between FRDM 
and DZ31 translates into a progressive filling of the through at $A\approx184$. 
Similarly, the erosion of the FRDM $S_{2n}$ saddle point around $Z/N=50/94$, corresponding to 
a smoothening in $\Delta_{2n}$ shown in \subrf{fig:mix_masses}{d}, produces the increase 
of the abundances at $A=144$.

It is worth noticing that the
contribution from the shell effects to the two-neutron separation energies is
usually subdominant compared to the bulk LDM part, as shown in
\subrf{fig:mix_masses}{b}. Despite this, the shaping of the \rpa\ abundances is
mostly driven by these relatively small local changes, which impact the trend
of neutron separation energies and the magnitude of neutron shell gaps.

By combining Fig.~\ref{fig:YZYA_mix} with Table~\ref{tab:rms_mix}, we
conclude that changes in nuclear masses related to shell effects of the order
of 570~keV translate to changes in the final abundance distribution up to a
factor of two. However, it is important to note that the impact on the abundances is not
necessarily proportional to the magnitude of the variations in the shell effects. The reason is
that variations in shell effects become more relevant when they induce
structural changes in the behavior of neutron separation energies, such as the
appearance or disappearance of local saddle points~\cite{Arcones2011}. As a consequence,
the impact of changes in the shell effects strongly depends on the trends of the underlying
mass surface. 
To better quantify the impact of changes in local shell effects in
\rpa\ abundances, we compute the $\Delta Y$ average at 1~Gyr for the different 
mixed models. The results are summarized in Table~\ref{tab:rms_mix} for nuclei with $Z > 20$,
$A > 50$, $Y(Z)$, $Y(A) > 10^{-7}$. We find that, indeed, changes in local shell effects are 
quantitatively linked to changes in the abundances distribution: changing the shell 
effects by 0.57 keV (i.e., the $S_{2n}$ by 0.24~MeV and $\Delta_{2n}$ by 0.27~MeV) changes 
the mass abundances by approximately 35\% in a rather consistent fashion. The impact on 
the elemental abundances is not as constant, but still in a relatively narrow range 
between $22-30$\%.
It is important to note that the $\sigma_\textrm{FRDM}$ values presented in Table~\ref{tab:rms_mix} account only for changes in local shell effects. In most cases, we expect the $\sigma$ between mass models to be larger due to the additional change in the bulk term, as shown by the comparison between the 100/0 and DZ31 models.
Also, the results obtained for these two models provide additional evidence that the rms 
error of nuclear masses can be a misleading indicator for the precision required in \rpa\ studies.
Despite the large difference in rms errors between 100/0 and DZ31, both mass models predict very similar abundance distribution. 
Finally, we point out that the $\langle Y \rangle$ values presented in Table~\ref{tab:rms_mix}
could be affected by the impact of shell effects in nuclear processes such as $\beta$ decays and fission, which can introduce non-local effects through, e.g., beta-delayed emission of neutrons and fission fragments.

\section{Conclusions\label{sec:conclusions}}
We have shown that variations in predicted nuclear masses do not necessarily
correlate with changes in the \rpa\ abundance distributions.  We employ the LDM
to decompose the mass surface into a smooth contribution and local shell
effects for the two mass models considered: FRDM and DZ31.  By performing
detailed nuclear network calculations simulating the occurrence of \rpn\ in
neutron star mergers, we find that large changes in the bulk properties that
result in very different theoretical nuclear masses, have little impact on the
predicted elemental and mass abundances. The abundance distribution is instead
driven by local changes in the (two-)neutron separation energies arising from
nucleon shell effects, which can be identified by means of neutron shell-gap
energies. We assess the sensitivity of \rpa\ abundances to such local
variations by mixing the shell effects of FRDM and DZ31. The resulting mass
tables gradually transition from one model to the other, while keeping the same
bulk properties. The \rpa\ abundances predicted by the mixed mass tables 
show a correlated smooth transition, providing evidence that local changes
in mass surfaces are the main drivers of \rpa\ abundances, despite accounting
for only a minor fraction of the overall $S_{2n}$ values. While the impact of
variations in shell effects on the abundance distribution is not always
proportional to the magnitude of the mass changes,
and depends on the local $S_{2n}$ and $\Delta_{2n}$ trends of the underlying mass model,
we find that variations in
nuclear masses related to shell effects of the order of 570~keV translate to
an average change in the final mass abundance distribution of 35\%.

Our results suggest that theoretical
approaches should aim to determine how robust the local mass trends are against
variations of the nuclear interaction/functional and improvements in the
many-body approach (such as inclusion of beyond mean-field effects). In this
context, ab-initio calculations, that currently do not provide accurate values
of the masses but that may capture the local trends~\cite{Hergert2020}, may
still provide valuable information for \rpa\ studies. Similarly, experimental
measurements devoted to nuclear masses for \rpa\ nucleosynthesis should aim to
determine mass trends across extended regions, rather than
measure isolated individual nuclei.
Furthermore, fitting protocols of nuclear interactions and training processes
of machine-learning algorithms devoted to large-scale calculations of
neutron-rich nuclei should include experimental information of (two-)neutron
separation and shell gap energies. While a proper reproduction of nuclear
masses is a crucial challenge for nuclear theory, this work demonstrates that
the rms error between theoretical and experimental masses may not necessarily
determine which nuclear mass models are the best suited for \rpa\ studies. A
similar conclusion was reached in Ref.~\cite{Sobiczewski.Litvinov:2014}
comparing the predictive power of different mass models to masses which were
not experimentally known at the time of the model calibration.  Finally, we
make explicit that the method employed in this work is one possible means of
producing variations in nuclear masses that do not affect the abundance
distribution. The identification of additional means of variations, and their
connection to the properties of the underlying nuclear interaction, will open a
new avenue to better understand the physics shaping the \rpa\ abundance
pattern.

\begin{acknowledgements}
  We thank M.-R.~Wu, F.-K.~Thielemann, W.~Nazarewicz, and K.~Langanke for
  helpful discussion and comments. SG acknowledges support by the Spanish
  Agencia Estatal de Investigaci{\'o}n (AEI) of the Ministry of Science and
  Innovation (MCIN) under grant agreements No.~PID2021-127890NB-I00 and
  No.~RYC2021-031880-I funded by MCIN/AEI/10.13039/501100011033 and the
  ``European Union NextGenerationEU/PRTR''.  This work benefited from support
  by the U.S. Department of Energy, Office of Science, Office of Nuclear
  Physics, under Award Number DE-SC0023128 (CeNAM).  GMP acknowledges support by
  the European Research Council (ERC) under the European Union’s Horizon 2020
  research and innovation programme (ERC Advanced Grant KILONOVA No. 885281).
  GMP and AB acknowledge support by the Deutsche Forschungsgemeinschaft (DFG,
  German Research Foundation) through Project - ID 279384907 - SFB 1245
  (subprojects B01, B06, B07), and the State of Hesse within the Cluster
  Project ELEMENTS\@. AB and VV acknowledge support by the European Union (ERC,
  HEAVYMETAL, 101071865).
\end{acknowledgements}

\bibliography{library.bib}

\end{document}